\documentclass{aa}

\usepackage{graphics}

\def\iue{\mbox{\it IUE}} 
\def\ines{\mbox{\it INES}} 
\def\newsips{\mbox{\it NEWSIPS}} 
\def\iuesips{\mbox{\it IUESIPS}} 
 
\begin{document} 

\thesaurus{23(03.13.2;03.20.1;13.21.2)}

\title{The INES System III: Evaluation of IUE NEWSIPS High Resolution Spectra}

\author{
R. Gonz\'alez-Riestra\inst{1}\fnmsep\thanks{\it{Previously:} ESA--IUE Observatory}
\and A. Cassatella\inst{2,3}\fnmsep $^*$
\and E. Solano\inst{1}\fnmsep $^*$ 
\and A. Altamore\inst{3}
\and W. Wamsteker\inst{4}\fnmsep\thanks{Affiliated to the Astrophysics 
Division, SSD, ESTEC}}

\offprints{R. Gonz\'alez-Riestra}
\mail{ch@laeff.esa.es}

\institute
{
Laboratorio de Astrof\'{\i}sica Espacial y F\'{\i}sica Fundamental,
VILSPA, P.O. Box 50727, 28080 Madrid, Spain 
\and
Istituto di Astrofisica Spaziale, CNR, Area di Ricerca Tor Vergata,
Via del Fosso del Cavaliere, 00133 Roma, Italy 
\and
Dipartimento di Fisica E. Amaldi, Universita degli Studi di Roma Tre, Via
della Vasca Navale 84, 00146 Roma, Italy
\and
ESA-IUE Observatory, VILSPA, P.O. Box 50727, 28080 Madrid, Spain }

\date{Received / Accepted}

\authorrunning{R. Gonz\'alez-Riestra  et al.}

\titlerunning{Evaluation of IUE High Resolution Spectra}

\maketitle

\begin{abstract}
 
This paper discusses the overall quality of \iue\ high
resolution data processed with the \newsips\ software in terms
of flux and wavelength accuracy.   It also describes the
processing of \newsips\ high resolution spectra within the
framework of the ESA ``IUE Newly Extracted Spectra'' (\ines)
System. This system provides the \iue\ high resolution data in
two formats: the high resolution ``concatenated'' spectra, in
which the spectral orders are connected, eliminating the overlap
regions through a procedure designed to optimize the
signal-to-noise ratio at the edges of the orders, and the
``rebinned'' spectra, i.e. the high resolution concatenated
spectra resampled into the low resolution wavelength domain. Our
study reveals the existence of a significant discrepancy in the
wavelength scales of short and long wavelength \newsips\ high
resolution spectra. The \ines\ processing applies a correction
of +17.7 km~s$^{-1}$ to the wavelength scale of the high
resolution SWP spectra in order to provide an internally
consistent velocity scale. Similarly, suitable corrections have 
been applied to long wavelength small aperture spectra. The new
wavelength scale is, within the errors, in good agreement with
the optical velocity scale.


\end{abstract}

\section{Introduction}

\label{sec:intro}

The \iue\ \newsips\ processing system was developed with the aim
of creating a ``Final Archive'' of \iue\ data to be made
available to the astronomical community as a legacy after the
end of the project. This archive would include all the \iue\
spectra re-processed with improved algorithms and up-to-date
calibrations. The \newsips\ processing system is fully described
by Nichols and Linsky (1996) and Nichols (1998). Technical
details are given in the \newsips\ Manual (Garhart et al. 1997).
The introduction of new techniques to perform the geometric and
photometric corrections led to a substantial improvement in the
signal-to-noise ratio of the final spectra.  Further
improvements in the quality of the high resolution data arise
from the new method to determine the image background and from
the improved ripple correction and absolute calibration
(Cassatella et al. 1999, hereinafter Paper II). The background
subtraction has always been one of the more critical issues in
the processing of \iue\ high resolution spectra, particularly at
the shortest wavelengths, where orders crowd and an accurate
estimate of the background is essential for a correct flux
extraction. This problem has been overcome in \newsips\ through
the derivation of a bi-dimensional background (see Smith 1999
for a description of the method).

The goal of the \ines\ processing system was to correct the
deficiencies found during the scientific evaluation of the data
processed with \newsips\ for the \iue\  Final Archive, and to
provide the output data to the users in a simple way requiring a
minimum knowledge of the operational and instrumental
characteristics of \iue\ (Wamsteker et al. 1999). The
modifications introduced in the processing of low resolution
data have been described by Rodr\'{\i}guez-Pascual et al. (1999,
Paper I). As for  high resolution data, the \ines\ system
provides two output products derived from the \newsips\ MXHI
(i.e. high resolution extracted spectra) files: the
``concatenated'' spectrum, where the spectral orders are merged
eliminating the overlap regions, and the ``rebinned'' spectrum,
which is the concatenated spectrum resampled to the low
resolution wavelength step. Both concatenated and rebinned
spectra include an error vector calibrated in absolute flux
units.  The inconsistency between the high resolution short and
long wavelength scales in \newsips\ has been  corrected for in 
the \ines\ concatenated spectra.

\begin{table*}

\caption{Radial velocities obtained from \newsips\ high resolution spectra}
\begin{center}
\begin{tabular}{l l l l}

\hline
Target		&	SWP			&	LWP			&	LWR		\\
\hline
RR Tel		& -69.5$\pm$6.5 [106] em. lin.	   &-49.3$\pm$3.0 [170] em. lin.   & -51.0$\pm$4.4 [132] em. lin.	\\     
$\zeta$ Oph	& -24.7$\pm$3.9 [60] SII, SiII	   &-13.4$\pm$2.6 [24] FeII,MnII   & -10.2$\pm$2.5 [69] FeII, MnII \\
BD+28~4211	& -22.8$\pm$6.4 [130] SII,SiII,CII & -5.3$\pm$3.6 [40] MgII K      &  -1.1$\pm$4.8 [21] MgII K and H, MgI \\
HD~60753	&  18.8$\pm$6.6 [62] SII,OI,SiII   & 32.4$\pm$6.9 [70] MgII K	   &  29.5$\pm$5.8 [24] MgII K and H, MgI \\
HD~93521	& -38.8$\pm$4.0 [68] SII,OI,SiII   &-20.7$\pm$6.9 [6]  MgII K	   & -19.9$\pm$5.4 [18] MgII K and H \\
BD+75~325	& -16.4$\pm$4.6 [121] SII,SiII,CII &  6.2$\pm$2.2 [43] MgI	   &   6.6$\pm$3.1 [33] MgII K	\\
$\lambda$ Lep	&				   & 21.2$\pm$3.9 [37] MgII K, MgI &  19.5$\pm$4.9 [44] MgII K and H  \\
$\zeta$ Cas	&				   & -1.1$\pm$4.2 [9]  MgII K	   &   1.5$\pm$4.4 [72] MgII K and H \\
$\eta$ UMa	&				   & -1.1$\pm$3.1 [17] MgII K	   &   3.1$\pm$7.1 [82] MgII K and H \\
\hline
\end{tabular}
\end{center}
\vspace{0.1cm}
\noindent The entries for each star are: mean radial velocity, rms (km~s$^{-1}$), number of
measurements, and lines used (em.lin.: emission lines).\\
\noindent All measurements refer to the large aperture.\\
\label{tab:rvel}
\end{table*}

In the first part of this paper we discuss the overall quality
of \newsips\ high resolution spectra in terms of accuracy,
stability and repeatability of wavelength and flux measurements
(Section \ref{sec:dataev}).  The second part deals with the
\ines\ processing of high resolution data, describing the order
concatenation and rebinning procedures (Sections
\ref{sec:concat} and \ref{sec:rebin}, respectively). Finally,
the application of the correction to the wavelength scale is
discussed (Section \ref{sec:wave}).

\section{NEWSIPS Data quality evaluation}

\label{sec:dataev}

The overall quality of \iue\ high resolution spectra processed
with \newsips\ has been evaluated by studying the accuracy and
the stability of wavelength determinations, the accuracy of
equivalent width measurements, the flux repeatability and the
residual camera non--linearities.

The analysis is based on a large number of spectra, mainly of
the \iue\ standard stars. The spectra have been corrected for
the echelle blaze function and calibrated in terms of absolute
fluxes according to the procedure described in Paper II. 
Hereinafter, wavelengths are assumed to be in the heliocentric
reference frame and in vacuum.

\subsection{Wavelength accuracy}

To assess the wavelength accuracy three aspects have been
considered separately:

a) the accuracy and repeatability of wavelength measurements of
a given spectral feature in several spectra of the same star,

b) the stability of the wavelength scale along the full spectral
range,

c) the consistency of radial velocity determinations obtained
from the SWP, LWP and LWR cameras.

\subsubsection{Expected accuracy}

One of the most important limitations to the wavelength accuracy
of \iue\ high resolution spectra are the target acquisition
errors at the nominal center of the spectrographs entrance
apertures. These, if large enough, can also affect the quality
of the ripple correction (see Paper II).

From the \newsips\ dispersion constants and the central
wavelengths of the spectral orders it can be readily deduced
that the velocity dispersion corresponding to one pixel on the
image is practically constant all through the range covered by
the cameras, and namely:\\

$\Delta$V=7.73 $\pm$ 0.05 km~s$^{-1}$ for SWP 

$\Delta$V=7.26 $\pm$ 0.09 km~s$^{-1}$ for LWP

$\Delta$V=7.26 $\pm$ 0.03 km~s$^{-1}$ for LWR\\

Taking into account that the plate scales are 1.530, 1.564 and
1.553 arcsec/pix for SWP, LWP and LWR, respectively (Garhart et
al. 1997), an acquisition error of 1 arcsec along the high
resolution dispersion direction would lead to a constant
velocity offset of 5.1 km~s$^{-1}$ for SWP, 4.6 km~s$^{-1}$ for
LWP and 4.7 km~s$^{-1}$ for LWR. Since the pointing/tracking
accuracy is usually better than 1 arcsec, we can consider 5
km~s$^{-1}$ as a reasonable  upper limit to the expected
wavelength accuracy. Wavelength errors substantially larger
might arise internally in the data extraction procedures.

\subsubsection{Repeatability of wavelength determinations}

To obtain a reliable information on the self-consistency of
wavelength determinations, we have attempted to reduce the
effects of spectral noise by performing a large number of
measurements of selected narrow and symmetric absorption lines
from the interstellar medium, which are strong in some of the
\iue\ calibration stars, as well as of many emission lines in RR
Tel.  Out of the \iue\ standards, we have selected BD+28~4211,
HD~60753, HD~93521, BD+75~325, $\lambda$ Lep (HD~34816), $\zeta$
Cas (HD~3360) and $\eta$ UMa (HD~120315). In addition, we have
also used spectra of the star $\zeta$ Oph (HD~149757). The
present measurements refer only to large aperture spectra.

The interstellar lines selected for the SWP range were: SII
1259.520~\AA, SiII 1260.412~\AA, OI 1302.168~\AA, SiII
1304.372~\AA\ and CII 1334.532~\AA, 1335.703~\AA. For the long
wavelength range we have used the K and H components of the MgII
doublet at 2796.325~\AA\ and 2803.530~\AA\ and MgI 2852.965~\AA.
In the case of $\zeta$ Oph we have also measured MnII
2576.877~\AA, 2594.507~\AA\ and several FeII lines.    The cases
in which the distortion of the profile  due to close-by reseau
marks (in particular for OI 1302 and CII 1334) precluded the 
accurate determination of the line position have not been taken
into account in the final statistics.  Laboratory wavelengths
have been taken form Morton (1991).

The mean values of the radial velocities, the corresponding rms
deviation and the number of independent measurements are
reported in Table \ref{tab:rvel} for each target and camera.
According to this table, the rms repeatability error on radial
velocities, averaged over the three cameras is 4.6$\pm$1.5
km~s$^{-1}$. This value is smaller than the upper limit of about
5 km~s$^{-1}$ expected from acquisition errors.  Considering the
presence of spectral noise, we can safely conclude that the
repeatability of wavelength (or velocity) determinations is
satisfactory.

The results in Table \ref{tab:rvel} indicate that the radial
velocities derived from the two long wavelength cameras are
consistent while, on the contrary, the velocities derived from
SWP spectra are systematically more negative.  This, and other
considerations about the consistency of radial velocity
determinations from the three cameras will be discussed in
Section \ref{sec:wave}.

\subsubsection{Stability of the wavelength scale along the full spectral range}

We have studied the accuracy of the wavelength scale over a wide
spectral range to look for possible time-dependent distortions
across the camera faceplate. To this purpose, we have selected 6
SWP, 11 LWP and 3 LWR spectra of the emission line object RR Tel
obtained at different epochs. For each spectrum we have measured
the peak wavelengths of several emission lines chosen among
those reasonably well exposed and with the cleanest profiles,
covering the full spectral range. The highest excitation lines,
such as those from [MgV], were purposely excluded because they
provided systematically higher negative radial velocities
probably due to stratification effects within the nebular
region. The mean radial velocities of RR Tel are -69.5$\pm$6.5
km~s$^{-1}$, (SWP), -49.3$\pm$3.0 km~s$^{-1}$ (LWP), and
-51.0$\pm$4.1 km~s$^{-1}$ (LWR). The total number of
measurements are 106, 170 and 132 for SWP, LWP and LWR,
respectively.  Since the errors are of the same order as the
repeatability errors quoted in the previous section, we conclude
that the wavelength scales do no present appreciable distortions
over the  wavelength range covered and, within the observational
errors, are stable over the period of time considered
(1983-1994, 1985-1995 and 1978-1983 for SWP, LWP and LWR,
respectively).

\subsubsection{Radial velocity determinations from the Mg~II doublet}

The present analysis has revealed the existence of an
inconsistency in the radial velocities derived from the MgII K
(2796.32 \AA) and H (2803.53 \AA) lines as measured in the LWP
camera, where the two lines are present in both orders 82 and
83.   To quantify this discrepancy, we have measured the
velocities of the Mg~II interstellar lines in 89 spectra of
five  \iue\ standard stars.

 We find that, in the LWP camera, the radial velocity difference
V$_{\rm H}$-V$_{\rm K}$ is  -10.8$\pm$2.5 km~s$^{-1}$ when
measured in order 83, and  11.7$\pm$1.5 km~s$^{-1}$ when
measured in order 82. We find also that there is a discrepancy
in the velocity of the K line measured in the two orders
(V$_{\rm 83}$-V$_{\rm 82}$=-20.5$\pm$0.9 km~s$^{-1}$), while  
the velocities of the H line measured in the two orders are
consistent within 2 km~s$^{-1}$, on average. Since the velocity
derived from  other interstellar lines (e.g.  MgI 2852.965~\AA) 
is fully consistent with the measurements from the K line in 
order 83, we conclude that  only this line provides correct
radial velocity values. In \ines\ concatenated spectra (Section
\ref{sec:concat}) the K line comes from order 83 and the H line
from order 82.  Therefore, there is  a systematic difference
V$_{\rm K}$-V$_{\rm H}$=-8.8$\pm$1.3 km~s$^{-1}$  between the
velocities determined from the two lines,  being the correct
value only that  given  by the K line. 

A similar study performed in LWR spectra, where the K line
appears only in order 83, shows that this problem is not present
in this camera, where the two Mg~II lines provide consistent
velocity values:  V$_{\rm K}$(order 83)-V$_{\rm H}$(order
82)=-1.0$\pm$1.2 km~s$^{-1}$.

\begin{figure} 
\resizebox{\hsize}{!} 
{\includegraphics{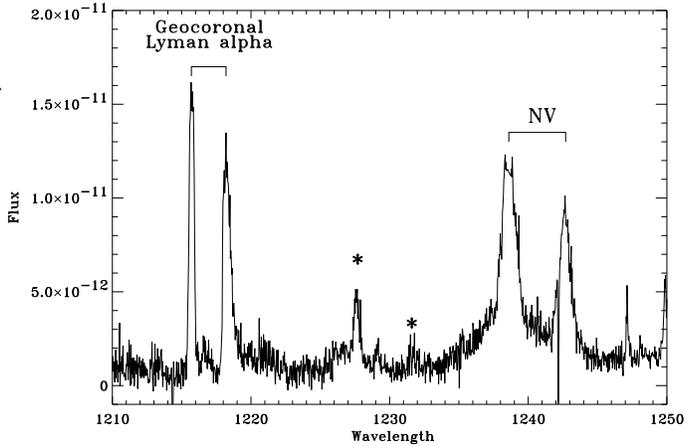}}  
\caption{A portion of the longest
available SWP exposure of RR Tel (SWP20246:  820 minutes). The
two Lyman $\alpha$ lines arise from geocoronal emission filling 
the large and the small apertures. The emission features labeled
with an asterisk are due to overspilling of the strong NV
emission into the adjacent order.}  
\label{fig:swprr} 
\end{figure}

\subsection{Background extraction}

It has been repeatedly pointed out that the background
extraction for high resolution spectra processed with \iuesips\
was not accurate enough especially shortward of 1400 \AA\ in the
SWP camera and 2400 \AA\ in the long wavelength cameras, as
denoted by the negative fluxes assigned to the wings of the
strongest emission lines and to the core of the saturated
absorption lines. As shown below, this effect is not present
anymore in spectra processed with \newsips, which makes use of
an upgraded background determination procedure  (Smith 1999).

Overestimating or underestimating the background level leads to
underestimating or overestimating the fluxes of the emission
lines and the equivalent widths of the absorption lines. In the
following, we report the tests done on the accuracy of the
equivalent widths to verify the correctness of the background
extraction. In addition, we have used the repeatability of the
equivalent widths determinations as an indirect test of the
stability of the background levels.

\subsubsection{SWP}

Figure \ref{fig:swprr} shows the profiles of the NV doublet
emission and of the broad Lyman $\alpha$ feature in the longest
exposure available of RR Tel. It is clearly seen that the wings
of these lines are not assigned negative values. Particularly
interesting are the NV ``ghost'' lines marked with an asterisk,
which are still present in \newsips\ data, but sensibly fainter
than in the \iuesips\ spectra, most likely due to the optimized
extraction slit. The presence of such spurious lines has
recently been reported by Zuccolo et al. (1997) and ascribed to
overspilling of the strong NV doublet into adjacent orders.

\begin{figure}
\resizebox{\hsize}{!}
{\includegraphics{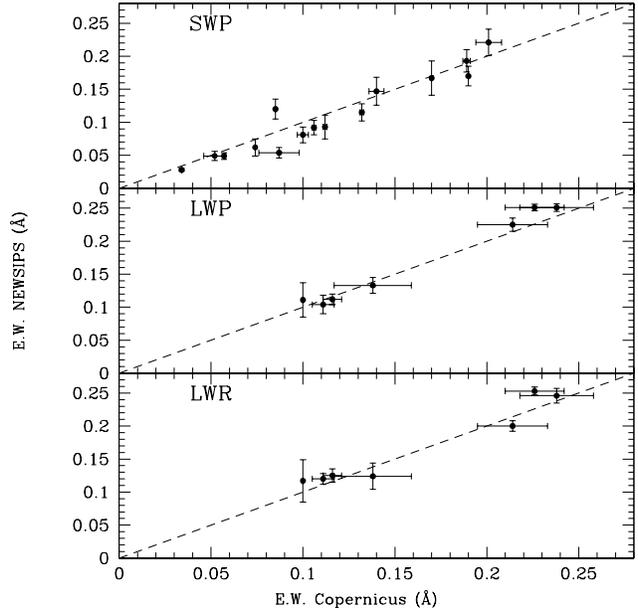}}
\caption{Comparison of the equivalent widths of interstellar
lines in the spectrum of $\zeta$ Oph as measured in Copernicus
and in \newsips\ spectra. Copernicus measurements have been
taken from Morton (1975). The dashed line in each panel 
indicates the 1:1 relation.}
\label{fig:morton}
\end{figure}

To verify the accuracy of the background subtraction, we have
compared the equivalent widths of the strongest interstellar
lines in four spectra of $\zeta$ Oph with those reported by
Morton (1975) obtained from Copernicus data. These latter
determinations are presumably not affected by background
determination problems, unlike the \iue\ echelle spectra near
the short wavelength end of the cameras. The results of the
comparison are given in  Fig.~\ref{fig:morton} and in Table
\ref{tab:swpew},  which provides  the mean and the standard
deviation of the four measurements. As it appears clearly from
the figure, the \newsips\ measurements are  consistent, within
the errors, with the values from Copernicus data, suggesting
that the background evaluation for SWP spectra is essentially
correct.

\begin{table}
\caption{Equivalent widths (in m\AA) in SWP spectra of $\zeta$ Oph compared with
Copernicus measurements}
\begin{center}
\begin{tabular}{l c c c }
\hline
Line		&	NEWSIPS			& Morton (1975)	\\
\hline
SII~1250.59	&	81$\pm$12		&	100$\pm$3	\\
SII~1253.81	&	92$\pm$11		&	106$\pm$2	\\
SII~1259.52	&	93$\pm$18		&	112$\pm$1	\\
SiII~1260.42	&	167$\pm$26		&	170	\\
CI~1277.24	&	62$\pm$13		&	74	\\
OI~1302.17	&	221$\pm$20		&	201$\pm$7	\\
SiII~1304.37	&	115$\pm$13		&	132$\pm$1	\\
CI~1328.83	&	49$\pm$7		&	52$\pm$6	\\
CII~1334.53	&	193$\pm$17		&	189$\pm$2	\\
CII~1335.70	&	147$\pm$21		&	140$\pm$4	\\
SiII~1526.71	&	170$\pm$15		&	190:	\\
FeII~1608.46	&	120$\pm$15		&	85:	\\
SiII~1808.01	&	54$\pm$8		&	87$\pm$11	\\
AlIII~1854.72   &	49$\pm$5		&	57:	\\
AlIII~1862.79   &	28$\pm$2		&	34:	\\
\hline
\end{tabular}
\end{center}
: means uncertain value
\label{tab:swpew}
\end{table}		

The stability of the background subtraction has been evaluated
by measuring the equivalent width of several strong interstellar
lines in a large sample of spectra of two standard stars. The
repeatability of the equivalent widths ranges from 10\%\ for the
strongest lines to 30\%\ for the faintest ones (Table
\ref{tab:ewrep}).

\begin{table}
\caption{Equivalent widths (in m\AA) in LWP and LWR spectra of $\zeta$ Oph compared with
Copernicus measurements}
\begin{center}
\begin{tabular}{l c c c}
\hline
Line		&	LWP		&	LWR	& Morton (1975)	\\
\hline		
ZnII~2026.16	&	111$\pm$26	& 117$\pm$32	&	100:	\\
FeII~2382.76 	&	251$\pm$6	& 246$\pm$11	&	238$\pm$20	\\
MnII~2576.88	&	133$\pm$12	& 124$\pm$20	&	138$\pm$21	\\
FeII~2586.65	&	225$\pm$10	& 200$\pm$8	&	214$\pm$19	\\
MnII~2594.51	&	112$\pm$8	& 125$\pm$10	&	116$\pm$5	\\
FeII~2600.17	&	251$\pm$5	& 253$\pm$6	&	226$\pm$16	\\
MnII~2606.48	&	104$\pm$14	& 120$\pm$8	&	111$\pm$6	\\
\hline
\end{tabular}
\end{center}
: means uncertain value
\label{tab:lwew}
\end{table}		

\begin{table}
\caption{Repeatability of equivalent widths measurements}
\centerline{SWP}
\vspace{0.3cm}
\begin{center}
\begin{tabular}{l c c }
\hline
Line	&	BD+28~4211	& BD+75~325	\\
\hline
SII~1250.59	&	32$\pm$9 (29)	&	34$\pm$8 (26)	\\
SII~1259.52	&	48$\pm$10 (28)	&	70$\pm$12 (35)	\\
SII~1260.42	&	92$\pm$12 (37)	&	128$\pm$14 (25)	\\ 
CII~1334.53	&	115$\pm$14 (36)	&	125$\pm$19	(35)	\\
\hline
\end{tabular}
\end{center}

\centerline{LWP}
\begin{center}
\begin{tabular}{l c c }
\hline
Line	&	BD+28~4211	& BD+75~325	\\
\hline
MgII~2796.53	&	257$\pm$20	(37)	&	369$\pm$50 (33)	\\		
MgI~2852.97	&	276$\pm$40	(11)	&	45$\pm$16 (10)	\\
\hline
\end{tabular}
\end{center}

\centerline{LWR}
\begin{center}
\begin{tabular}{l c c}
\hline
Line	&	$\eta$ UMa		& $\zeta$ Cas	\\
\hline
MgII~2796.35\AA	&	65$\pm$11 (41)	& 354$\pm$22 (36)	\\
MgII~2803.53\AA	&	75$\pm$23 (41)	& 345$\pm$23 (36)	\\
\hline
\end{tabular}
\end{center}
Equivalent widths in m\AA. The number of measurements is given in brackets. \\
\label{tab:ewrep}
\end{table}

\subsubsection{LWP}

\begin{figure}
\resizebox{\hsize}{!}
{\includegraphics{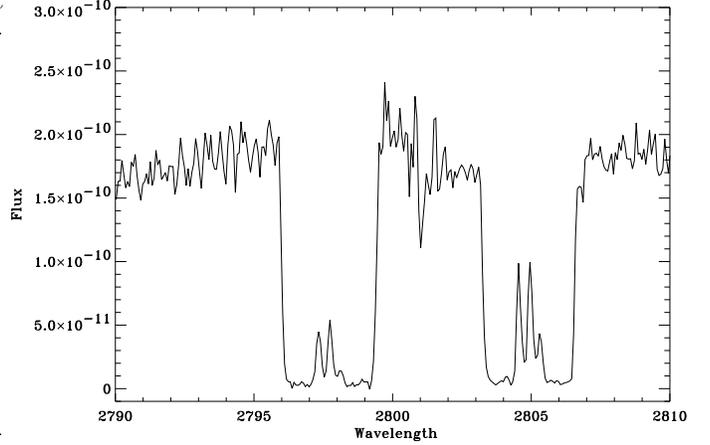}}
\resizebox{\hsize}{!}
{\includegraphics{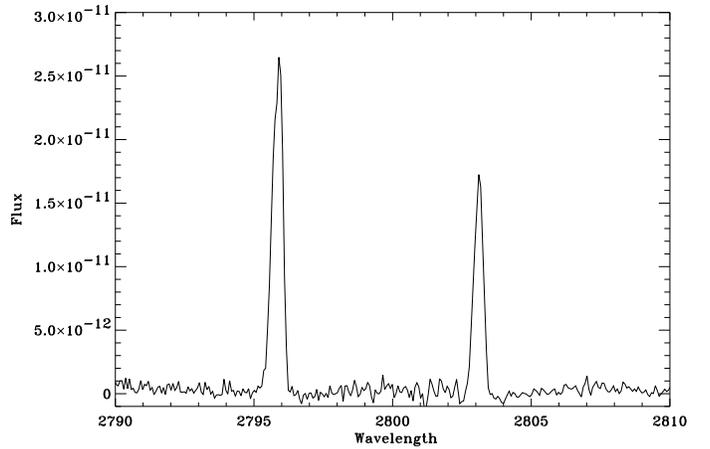}}
\caption{Top panel: The region of the MgII doublet in the
spectrum of SN 1987A  (LWP10194). Bottom panel: The same region
in the deepest LWP exposure of RR Tel (LWP25954).}
\label{fig:lwprr}
\end{figure}

We tested the accuracy of the background extraction from the
cores of strongly saturated absorption lines and  the wings of
strongly saturated emission lines. Fig.~\ref{fig:lwprr} shows a
portion of a spectrum of SN1987A centered around the Mg~II
doublet at 2800 \AA. It appears from the figure that the cores
of the absorption lines do not become systematically negative,
as expected for a correct background extraction. Shown in the
same figure is another example, that of the strongly saturated
Mg~II emission doublet in the longest LWP exposure available of
RR Tel (LWP25954): the lines wings do not reach negative fluxes,
as required.

A second test has been performed in four spectra of $\zeta$ Oph,
in which we have measured the equivalent widths of some strong
interstellar lines, and compared them with the Copernicus values
given by Morton (1975).  The results are summarized in
Fig.~\ref{fig:morton} and Table \ref{tab:lwew}, where are given
the mean and the standard deviation of the four spectra.

Finally, we have measured the equivalent widths of the MgII and
MgI lines in a large sample of spectra of BD+28~4211 and
BD+75~325. The repeatability errors range from 35\%\ for the
faint MgI line in BD+75~325, to less than than 10\%\ for the
strong lines. Results are shown in Table \ref{tab:ewrep}.

\begin{table}[htb!]
\caption{Flux repeatability of high resolution spectra}
\centerline{SWP}
\vspace{0.3cm}
\begin{center}
\begin{tabular}{l c c c}
\hline
Band		& Order	& set A	&	set B \\
\hline
1185--1190	& 116 	& 3.2	&	3.5	\\
1225--1230	& 112 	& 2.3	&	4.3	\\
1285--1290	& 107 	& 2.6	&	3.2	\\
1375--1380	& 100 	& 1.8	&	3.0	\\
1485--1490	& 93  	& 1.6	&	3.9	\\
1595--1600	& 86  	& 3.5	&	3.5	\\
1795--1800	& 77  	& 1.6	&	2.5	\\
\hline
\end{tabular}
\end{center}

Set A:  BD+28~4211: a) 4 spectra (Jun 86-Dec 86); b) 6 spectra (Nov 90-Jul 91), 
BD+75~325:  8 spectra (Dec 88-May 89),
HD~60753:   a) 4 spectra (Apr 80-May 80); b) 8 spectra (Feb 88-May 89),
HD~93521:   11 spectra (March 1987) 

Set B: 
BD+28~4211: 45 spectra (Dec 82- Aug 95),
BD+75~425:  53 spectra (Sep 85- Feb 95),
HD~60753:   38 spectra (Jul 79- Mar 95)

\vspace{0.3cm}
\centerline{LWP}
\begin{center}
\begin{tabular}{l c c c}
\hline
Band		& Order	& set A	& set B \\
\hline
2117--2122	& 109   & 4.9	& 6.3	\\
2457--2462	& 94  	& 2.0	& 3.5	\\
2922--2927	& 79  	& 1.8	& 2.7	\\
3117--3122	& 74  	& 3.0	& 4.0	\\
\hline
\end{tabular}
\end{center}

Set A: 
BD+75~325: a) 11 spectra (March 85-Dec 85); b) 9 spectra (Jan 89-Dec 89);
c) 8 spectra (Feb 92-Nov 92) \\
Set B: 
BD+28~4211: 34 spectra 	(Dec 82-Aug 95),
BD+75~325:  37 spectra 	(Sep 85-Feb 95),
HD~60753:   24 spectra 	(Jul 86-Aug 95)

\vspace{0.3cm}
\centerline{LWR}
\begin{center}
\begin{tabular}{l c c c}
\hline
Band		& Order	& Set A & Set B \\
\hline
2117--2122	& 109 	& 4.2	& 5.5 	\\
2457--2462	& 94  	& 3.4	& 6.3	\\
2922--2927	& 79  	& 1.8	& 4.3	\\
3077--3082	& 75  	& 2.4	& 5.0	\\
\hline
\end{tabular}
\end{center}
Set A: 
$\zeta$ Cas: a) 4 spectra (Dec 81-Dec 82); b) 5 spectra (Sep 83-Jul 84),
$\eta$ UMa: a) 6 spectra (Mar 81-Apr 82); b) 5 spectra (Aug 82-Jun 86) \\
Set B: 
$\zeta$ Cas: 30 spectra (Feb 81-Feb 87),
$\eta$ UMa:  42 spectra (Sep 78-Jul 90) \\ 
\label{tab:rep}
\end{table}

\subsubsection{LWR}

As for the LWR camera, we have verified  the accuracy of the
background subtraction by  measuring the equivalent widths of
six strong FeII and MnII interstellar lines in four spectra of
$\zeta$ Oph, and compared them with measurements based on
Copernicus data.  As shown in Fig.~\ref{fig:morton},  there is a
good agreement between the two sets of equivalent widths, and no
systematic departures are found.

The repeatability of equivalent width determinations has been
determined measuring the equivalent widths of the MgII doublet
in a large sample of spectra of $\eta$ UMa and $\zeta$ Cas and
$\lambda$\ Lep. The repeatability error ranges from 30\% for the
weak lines to less than 5\% for the strongest ones (Table
\ref{tab:ewrep}).

\subsection{Flux repeatability}

Two different tests have been performed to assess the flux
repeatability of  high resolution spectra. In the first one we
selected restricted samples of spectra obtained close in time,
and measured the ripple corrected net fluxes, i.e. without
applying the time sensitivity degradation and the temperature
dependence corrections. The second test has been performed on
larger samples covering extended periods of time, measuring the
absolutely calibrated fluxes, which include time and temperature
corrections. In all case we have averaged the flux over a narrow
wavelength interval free of lines. The flux repeatibility was
defined  as the percent rms deviation from the mean value. The
results are summarized in Table \ref{tab:rep}.

\subsubsection{SWP}

For the first test we have used 41 high resolution spectra of 
\iue\ calibration standards grouped into sets of data with a
similar  exposure level and obtained close enough in time.  The
test was done in six bands 5 \AA\ wide. In Table \ref{tab:rep}
(under ``set A'') we report the percent rms deviation. The
repeatability of spectra obtained sufficiently close in time is
about 2\%. The second test was made on a larger number of
spectra with similar exposure time, without restricting the date
of observation (``set B''). This test provides repeatability
errors ranging from 3 to 4\%. These somewhat larger errors are
due to the intrinsic uncertainties of the sensitivity
degradation correction algorithm.

\subsubsection{LWP}

The tests performed are similar to those described above for the
SWP camera.  The spectra in ``Set A'' consist of three groups
each containing images obtained in a restricted period of time. 
The flux repeatability was evaluated in four wavelength bands 5
\AA\ wide.  In this cameras, the repeatability errors can reach
the 5\%\ level near the short wavelength end of the camera, but
are a factor of two lower in the central bands. A similar test
performed on a larger set of spectra needing correction for the
time-dependent sensitivity degradation (``Set B'') provides
errors slightly larger, confirming that the sensitivity
degradation algorithm adopted for the LWP camera is essentially
correct.

\subsubsection{LWR}

 The flux repeatability was evaluated in four wavelength bands 5
\AA\ wide. The test performed on spectra  taken close in time,
``Set A'', shows that the repeatability errors reach 4\%\ near
the short wavelength end of the camera, decreasing in the region
of maximum sensitivity and increasing again at the longest
wavelengths.  In the ``Set B'' spectra the repeatability is
worse, reflecting the uncertainties in the time degradation
correction and also the instability of the camera after it
ceased to be routinely used.

\begin{table}[htb]
\caption{Linearity study for high resolution spectra}

\centerline{SWP}
\vspace{0.3cm}
\begin{center}
\begin{tabular}{l  c c c c c c }
\hline
Image	& t/t(opt)	& 1185\AA & 1285\AA & 1485\AA & 1785\AA \\
\hline
41467	& 0.50 	& 0.94	& 0.95	& 0.97  & 1.05	\\	
41346	& 0.68 	& 0.92	& 0.98  & 0.95	& 1.02  \\
42309	& 0.70	& 0.97	& 0.93	& 0.98	& 1.00	\\
41435	& 0.90	& 0.96	& 0.99	& 1.01	& 0.99	\\
41466	& 1.00	& 0.98	& 0.98	& 1.00	& 0.98  \\
42260	& 1.00	& 1.02	& 1.02	& 1.01	& 1.03	\\
41495	& 1.80	& 0.96	& 0.98	& 0.98	& 0.99	\\
\hline 
\end{tabular}
\end{center}

\centerline{LWP}
\begin{center}
\begin{tabular}{l c c c c c}
\hline
Group	& t/t(opt) & 2120\AA	& 2460\AA & 2925\AA & 3132\AA \\
\hline
a	& 0.27	& 1.03	& 1.03  & 0.98	& 0.98 \\ 
b	& 0.41	& 1.00	& 1.02  & 0.99	& 0.98 \\ 
c	& 0.67	& 0.99	& 1.02	& 1.03  & 1.00 \\
d	& 0.83	& 0.95	& 0.98	& 1.00	& 0.97 \\
e	& 1.00	& 1.00	& 1.00	& 1.00  & 1.00 \\
f	& 1.33	& 0.93	& 1.03	& 1.00	& 1.03 \\
g 	& 2.07	& 1.01	& 1.04	& 1.04	& 1.08 \\
\hline
\end{tabular}
\end{center}
a: 2 spectra of BD+28~4211 \\
b: 1 spectrum of BD+28~4211 \\
c: 2 spectra of BD+28~4211 \\
d: 2 spectra of BD+75~325 \\
e: reference group: 4 spectra of BD+75~325 and 4 spectra of BD+28~4211 \\
f: 1 spectrum of BD+75~325 \\
g: 1 spectrum of BD+75~325 \\

\centerline{LWR}
\begin{center}
\begin{tabular}{l c c c c c}
\hline
Image	& t/t(opt) & 2120\AA	& 2460\AA & 2925\AA & 3080\AA \\
\hline
9955	& 0.50	& 0.75 	& 0.98  & 0.99	& 0.91 \\ 
9954	& 1.00	& 1.00	& 1.00  & 1.00	& 1.00 \\ 
9113,9953	& 2.25	& 1.18	& 1.05	& 1.00  & 1.01 \\
8116	& 2.50	& 1.12	& 1.04	& 1.01	& 1.00 \\
\hline
\end{tabular}
\end{center}
\label{tab:lin}
\end{table}

\subsection{Flux Linearity}

Despite the linearity correction applied during the processing,
residual non linearities are still present in \iue\ data. This
effect has been evaluated in Paper I for low resolution data. In
what follows we discuss this effect in high resolution spectra.
The method followed consists on studying a set of spectra of the
same star with different exposure times obtained, whenever
possible, very close in time (preferably on the same observing
shift) and with similar camera temperatures. The variation of
the flux with the level of exposure (or the exposure time)
defines the flux linearity. Unfortunately there exist  few sets
of high resolution data suitable for this study, and a slightly
different approach has been taken here. The results are
summarized in Table \ref{tab:lin}.  In general, these results
are in good agreement with those derived in Paper I.

\subsubsection{SWP}

The most complete set of SWP spectra appropriate to assess the
linearity consists of seven images of the white dwarf
CD-38~10980 obtained in the period April-August 1991, with
exposure times ranging from 50\%\ to 180\%\ of the optimum value
(200 min). For these images we have measured the mean flux in
five 5 \AA\ wide bands. The fluxes in each band have been
averaged together and divided by the mean flux in the two 100\%\
exposures. The ratios so obtained indicate departures from
linearity ranging from -6\%\ at 1185 \AA\ to +4\%\ at 1785 \AA\
for the 50\%\ exposure and up to -5\%\ for the 70\%\ exposure.

\subsubsection{LWP}

There is not any complete set of high resolution data of the
same star which allows to study the LWP camera linearity. We
have constructed average spectra of different exposure levels of
the two standard stars BD+28~4211 and BD+75~325, and divided
them by the corresponding 100\% spectrum.  The exposure levels
covered range from 27\%\ to 207\%\ of the optimum exposure time.
The test was performed in four wavelength bands 5 \AA\ wide,
selected for being relatively free from strong absorption
lines.  The maximum departures from linearity, reaching 8\%, are
found for the 133\%\ level at 2120 \AA, and for the 207\%\ level
near the regions of maximum sensitivity of the camera. The
latter deviation can be easily understood in terms of
saturation.

\subsubsection{LWR}

The LWR high resolution linearity test has been performed with
five images of the standard star HD~93521 obtained in the period
July 1980 to February 1981, covering the range of exposure times
from 50\%\ to 250\%\ of the optimum value.  The test was
performed in four bands 5 \AA\ wide.  The maximum departures
form linearity (up to 25\%) are found at the shortest
wavelengths, where fluxes are underestimated by 25\%\ for the
50\%\ exposure and overestimated by 12\%\ for the 250\%\
exposure.

\section{INES Processing of High Dispersion Spectra}

\label{sec:inespro}

The starting point for the \ines\ processing of high resolution
data are the \newsips\ MXHI files.  The spectra have not been
re-extracted from the  bi-dimensional  files, as in the case
of the low resolution data (see Paper I). The \ines\ system
provides two output spectra for each high dispersion image:  the
``concatenated'' and the ``rebinned'' spectra. Both  include a
modified wavelength scale and the error vector calibrated in
absolute flux units. All these aspects are discussed in the
following sections.

\subsection{The high resolution concatenated spectra}

\label{sec:concat}

The main features of the \ines\ concatenated spectra are:

a) The overlap regions between adjacent orders are suppressed in
such a way that the less noisy portion of the orders is
retained.

b) The error vector is calibrated in absolute flux units.

c) The wavelength scale is modified to make consistent the
radial velocities obtained from the short and the long
wavelength cameras. The wavelength sampling of the MXHI files
has been retained.

d) The spectra are provided as FITS tables having the same
format as the low resolution spectra, i.e. they contain only
four columns: wavelength, absolute flux, error and quality
factor. This setting reduces significantly the download time for
remote data retrieval, and simplifies considerably the structure
of the \newsips\ MXHI files, which contain additional
information, not relevant for most investigations, such as the
position of the orders on the bi-dimensional image, the height
of the extraction slit and the order number for each extracted
point.

The procedure followed to concatenate the spectral orders is described in detail in the next paragraphs.

\begin{figure}
\resizebox{\hsize}{!}
{\includegraphics{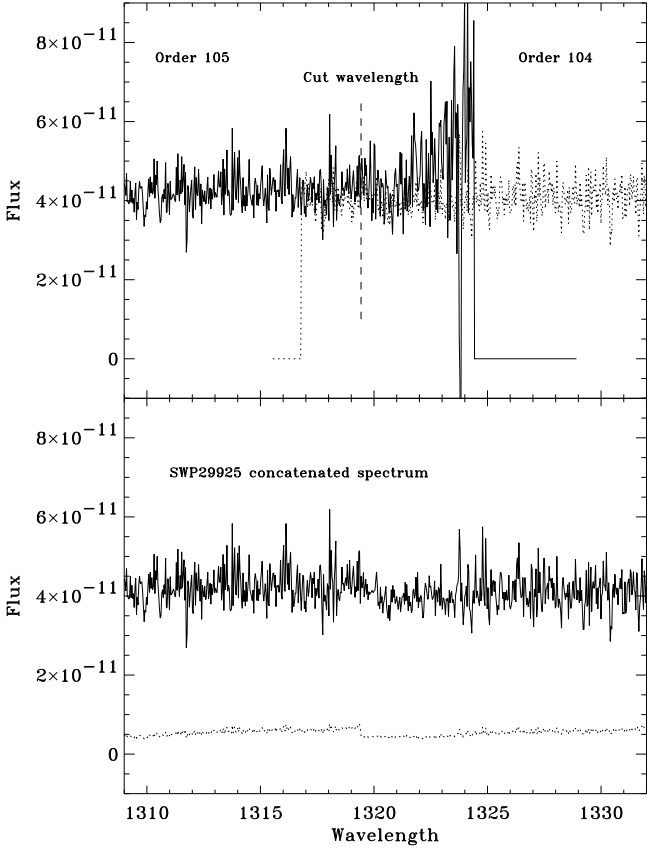}}
\caption{Example of the concatenation procedure for the SWP
camera. The top panel shows the individual orders from the MXHI
file.  The bottom panel shows the final concatenated spectrum
and the calibrated errors (dotted line).}
\label{fig:conswp}
\end{figure}  

\subsubsection{The concatenation procedure}

The most critical point in the concatenation of adjacent echelle
orders  is a suitable definition of the "cut wavelengths" so
that only the highest  quality data points of the overlap region
are retained.

\begin{figure}
\resizebox{\hsize}{!}
{\includegraphics{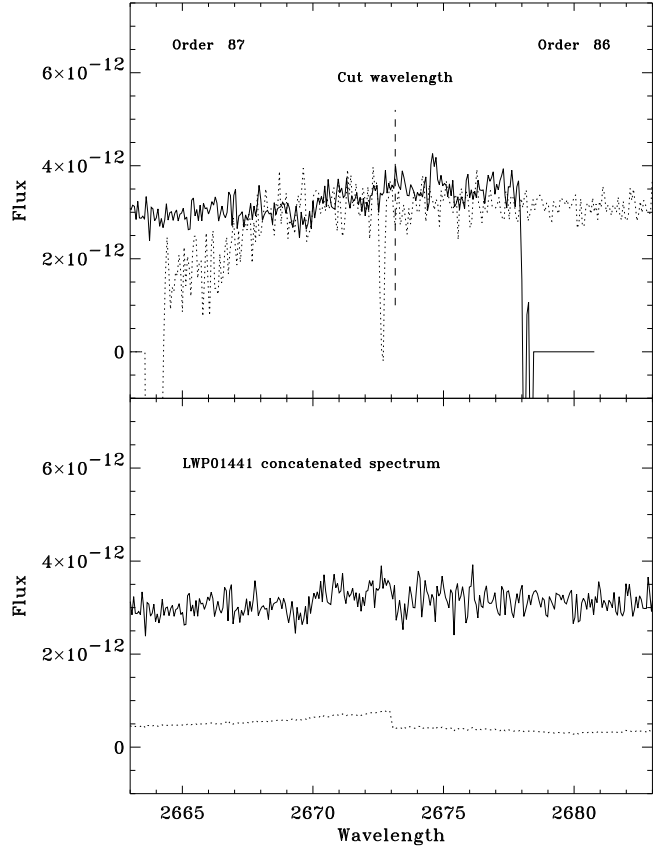}}
\caption{As figure \ref{fig:conswp} for the LWP camera.}
\label{fig:conlwp}
\end{figure}  

\begin{figure}
\resizebox{\hsize}{!}
{\includegraphics{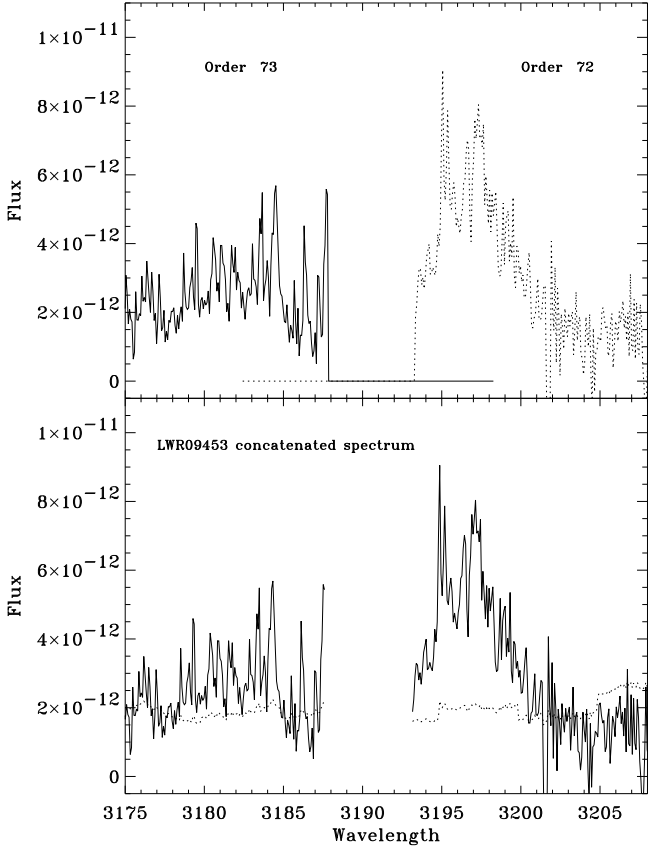}}
\caption{As Figure \ref{fig:conswp} for the LWR camera. 
In the spectral region shown orders do not overlap. The \ines\ concatenated
spectrum has no data in that region.}
\label{fig:conlwr}
\end{figure}  

In \iue\ high resolution spectra, the signal-to-noise level at
the edge of the orders is  different in the short and long
wavelength cameras. In the LWP and LWR cameras the
signal-to-noise is always lower at the short wavelength end of
the orders (except in the highest orders, see below). The
contrary happens in the SWP camera, where the long wavelength
edge of the orders is much noisier than the short wavelength
one.  Taking this into account, the cut wavelengths have been
defined as follows: \\

\noindent SWP:

\begin{equation}  
\lambda_{cut} = \lambda_{start} + (\lambda_{end}-\lambda_{start})/3.
\end{equation}

\noindent  LWP/LWR:
   
\begin{equation}  
\lambda_{cut} = \lambda_{start} + 2\times(\lambda_{end}-\lambda_{start})/3.
\end{equation}
where\\
$\lambda_{cut}$: cut wavelength between orders {\it m} and {\it m-1}\\
$\lambda_{start}$: start wavelength of overlap region (order {\it m-1})\\ 
$\lambda_{end}$: end wavelength of overlap region (order {\it m}) \\

These expressions are valid for all spectral orders except order
125 in the LWP camera and orders 120 to 125 in the LWR camera,
where the S/N ratio in order {\it m} is systematically higher
than  in order {\it m-1} in the overlap region. In these cases
only the points of order {\it m} are taken.

The above defined cut wavelengths (i.e. end wavelengths of order
{\it m}) can be computed as a function of order number as:

\begin{equation}
\lambda_{cut}(m) = A+\frac{B}{m}+\frac{C}{m^2}
\end{equation}

\noindent with the values of  A, B and C  given in Table \ref{tab:conc}.

For non overlapping orders (lower than 73, 77 and 76 for SWP,
LWP and LWR, respectively) only photometrically corrected pixels
have been included in the concatenated spectra (see
Fig.~\ref{fig:conlwr}).  The concatenated spectra cover the same
spectral range as the \ines\ low resolution spectra, i.e.
1150-1980 \AA\ for the SWP camera, and 1850-3350 \AA\ for LWP
and LWR.

Figures \ref{fig:conswp}, \ref{fig:conlwp} and \ref{fig:conlwr} show
examples of the concatenation procedure for the three cameras.

\begin{table}
\begin{center}
\caption{Parameters defining the cut wavelengths for the order concatenation}
\begin{tabular}{l c c c c }
\hline
Camera	& 	A	 &	   B	        &	C	& Orders \\
\hline
SWP LAP		&	24.3952	 &	132875.4838	& 325840.9715 &	120-73	\\
~~~~~~~~SAP	&	22.2095	 &	133293.4862	& 300351.2209 & 120-73 \\
LWP LAP 	&	-7.9697	 &	233257.6280	& 0           & 124-77	\\
~~~~~~~~SAP 	&	-7.7959	 &	233382.6450	& 0           & 124-77 	\\
LWR LAP 	&	-11.3459 &	233737.5903	& 0           & 119-76	\\	
~~~~~~~~SAP 	&	-11.2214 &	233876.9950	& 0           & 119-76 	\\
\hline
\end{tabular}
\label{tab:conc}
\end{center}
\end{table}

\subsubsection{The error vector}

The \newsips\ processing provides an error vector for the high
resolution spectra which is computed simply as the sum along the
extraction slit of the noise values for the individual pixels,
as derived from the camera noise model. Unlike the ``sigma'' of
the low resolution data, the ``sigma'' vector  in the MXHI files
is not flux calibrated but given in FN (Flux Number) units.

In the \ines\ high resolution data, the ``sigma'' spectrum is
provided in absolute flux units. The calibration is performed by
applying to the MXHI error vector the high resolution
calibration and the time sensitivity degradation correction.

\subsection{The resampled spectra}

\label{sec:rebin}

\begin{figure*} 
\begin{center}
\rotatebox{90} 
{\resizebox{!}{17cm}
{\includegraphics{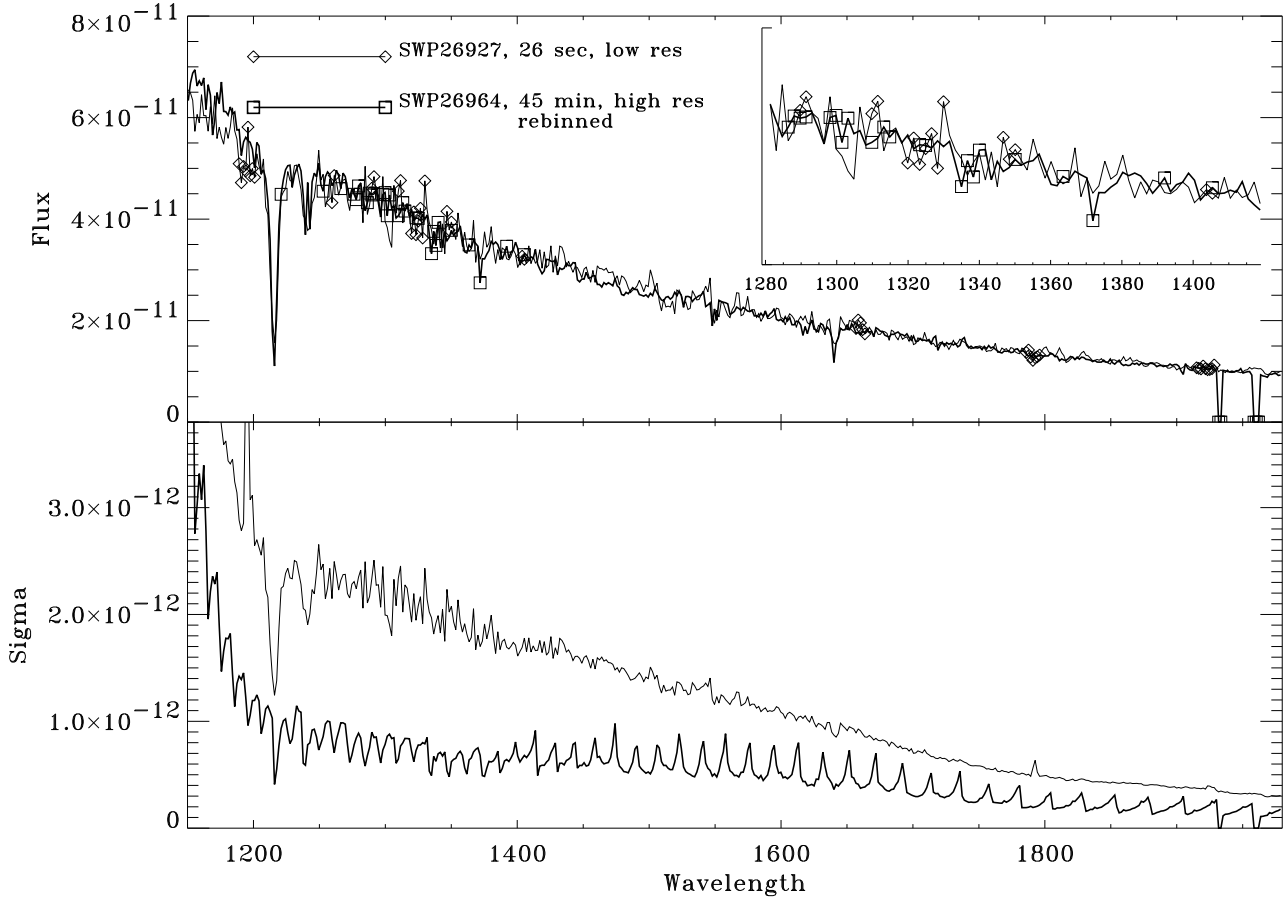}}} 
\caption{Comparison of low resolution (thin line) and high
resolution rebinned (thick line) spectra of the standard star
BD+28~4211. The top panel shows the flux spectra. Flagged pixels
are marked with diamonds (low resolution) and squares
(rebinned). The two gaps in the high resolution spectra longward
1900 \AA\ correspond to the regions where spectral orders do not
overlap, which have been assigned zero flux. The inset in the
top panel shows in more detail the region 1300--1400\AA. The
error spectrum is shown in the bottom panel (thin line: low
resolution, thick line: rebinned). The characteristic pattern of
the errors of the rebinned spectrum is due to the lower
signal-to-noise ratio at the edges of the individual echelle
orders.} 
\label{fig:rebin}
\end{center}
\end{figure*}

\begin{figure}
\resizebox{\hsize}{!}
{\includegraphics{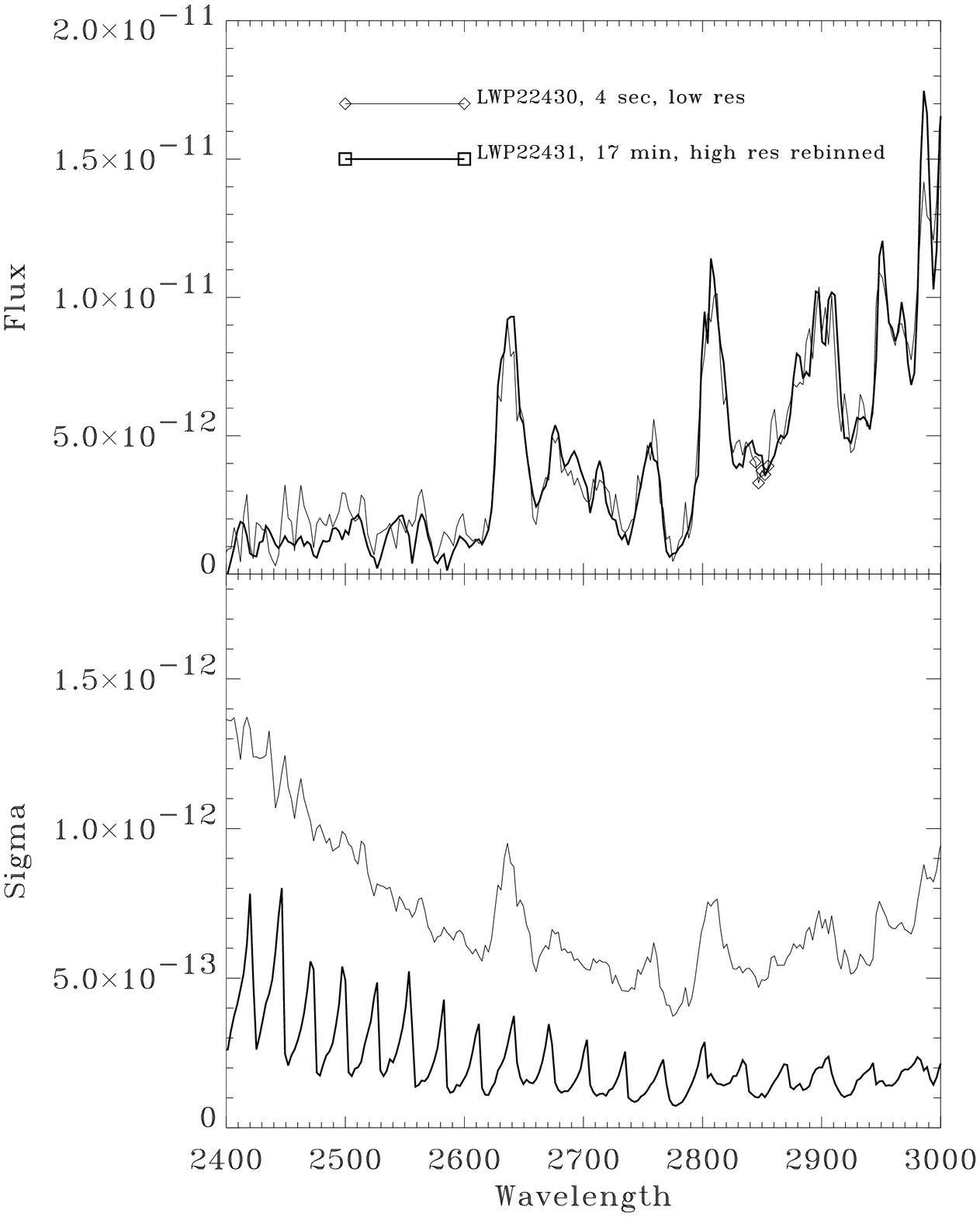}}
\caption{Similar to Figure \ref{fig:rebin} but comparing two LWP spectra of
Nova Cygni 1992.  The errors are shown in the bottom panel.}
\label{fig:rebinl}
\end{figure}

In the \ines\ Archive, each high resolution image has an
associated "rebinned" spectrum, which is obtained by rebinning
the  "concatenated"  spectrum at the same wavelength step size
as low resolution data. This  data set represents an important
complement to the low resolution archive,  and it is especially
useful for time variability studies. The {\it rebinned} spectra
have not been convolved with the low resolution Point Spread
Function, and  therefore have a better spectral resolution than
low dispersion spectra. Examples of rebinned spectra are shown
in Figures \ref{fig:rebin} and \ref{fig:rebinl}.

The high resolution {\it concatenated} spectra (derived as
described in the previous Section) are resampled into the low
resolution wavelength space following the procedure detailed
below.

\subsubsection{The rebinning procedure}

The concatenated spectra have been resampled into the \ines\ low
resolution wavelength domain as  defined in Paper I. The
sampling interval is 1.6764 \AA/pixel and 2.6693 \AA/pixel, for
the short wavelength and the long wavelength ranges,
respectively, and the wavelength coverage is 1150-1980 \AA\ for
SWP and 1850-3350 \AA\ for LWP and LWR.

The resampling has been performed so that the total flux is
conserved, that is, if {\it n} pixels with fluxes {\it $f_{1}$,
$f_{2}$, $\cdots$, $f_{n}$} are rebinned into one, the total
flux in the bin is:

\begin{equation}
 F = \sum^{i=n}_{i=1} (\lambda_{i}-\lambda_{i-1})(f_{i}+f_{i-1})/2.
\end{equation}

\noindent where the flux at the bin edges ({\it i=1, i=n}) is
calculated by linearly interpolating between the two adjacent
pixels. The flux of the final pixel is:

\begin{equation}
Flux = F/step
\end{equation}

\noindent being ``step'' the low resolution pixel size defined above.

\subsubsection{Errors}

The rebinned error spectrum is computed from the concatenated
error spectrum according to the following  expression:

\begin{equation}
E = \frac{\sqrt{\sum e_{i}^{2}}}{n}
\end{equation}

\noindent where  $e_i$ are the errors of the original pixels in
the concatenated spectrum.

\subsubsection{Flagging}

The quality flag assigned to each pixel in the resampled
spectrum is the sum of the flags of the original high resolution
pixels. Only the most relevant quality flags present in the high
resolution spectrum  have been transmitted to the rebinned
spectrum:

\begin{itemize}
\item[] -8192: 	Missing minor frames in extracted spectrum
\item[] -1024: 	Saturated pixel
\item[] -16: 	Microphonic noise (for the LWR camera only)
\item[] -8: 	Potential DMU corrupted pixel
\item[] -2:	Uncalibrated data point
\end{itemize}

Other flags (e.g. reseau marks) have not been taken into account
to avoid that a too large fraction of the pixels in the output
spectrum come out flagged with error conditions, despite their
quality is  not  significantly affected. Pixels corresponding to
the gaps between non-overlapping orders are flagged with ``-2''.

\section{The  Correction to the Wavelength Scale}

\label{sec:wave}

We already pointed out in Section \ref{sec:dataev} that there is
a significant discrepancy between the wavelength scales of short
and long wavelength spectra processed with \newsips. This
inconsistency, which is well above the accuracy of the
wavelength calibration, was already  present in \iuesips\ data 
(Nichols-Bohlin and Fesen 1986, 1990).

Table \ref{tab:rvel2} presents a summary of the radial
velocities of interstellar features in several stars, taken from
Table \ref{tab:rvel}, together with values from the literature
(v(lit)). The LWP and LWR  velocities of Table \ref{tab:rvel},
being consistent with each other, have been averaged together
into v(LW).

The literature value for RR Tel is from Tackeray (1977).  The
value quoted for $\zeta$ Oph  measured in the optical Ca~II K
and Na~I D lines has been  taken from Barlow et al. (1995). This
value is in good agreement with the velocities derived from GHRS
ultraviolet spectra by Savage et al. (1992), -14.9 km~s$^{-1}$,
and Brandt et al. (1996), -15.4 km~s$^{-1}$.  The velocities for
$\eta$ UMa and $\zeta$ Cas  correspond to measurements of the
Ca~II K and Na~I D optical lines reported by Vallerga et al.
(1993). The velocity quoted for $\lambda$ Lep  refers  to the
optical Ca~II doublet (Frisch et al. 1990), which presents two
components at 2 and 18 km~s$^{-1}$, which cannot be resolved
with \iue. The spectrum of HD~93521 presents up to nine
interstellar components, with the two strongest ones located at,
approximately, -10 and -60 km~s$^{-1}$ (Spitzer and Fitzpatrick
1993). In the IUE spectra all these systems appear blended, and
therefore, as in the case of $\lambda$ Lep, we cannot compare
reliably the IUE velocities with the optical values.

According to the data in Table \ref{tab:rvel2}, the mean
difference between long and short wavelength (large aperture)
velocities is 17.7 km~s$^{-1}$, i.e. SWP velocities are
systematically more negative.  The mean difference between the
long wavelength and the literature values is 8 km~s$^{-1}$.

A similar test was made to check the consistency between the
wavelength scale of spectra  taken through the large and small
apertures. Being the number of small aperture high resolution
spectra very limited,  useful data were available only for the
star  $\zeta$ Oph in the SWP and LWR cameras.  The wavelength
scales of the small and large aperture spectra  are fully
consistent in the short wavelength range: v(LAP)-v(SAP) =
1.1$\pm$6.7 km~s$^{-1}$, while for the LWR camera a significant
difference is found: v(LAP)-v(SAP) = -13.7$\pm$4.1 km~s$^{-1}$. 
The lack of a suitable data set precludes an accurate
determination of the offset between the  large  and small
aperture scales in LWP spectra, but the limited tests performed
seem to indicate that small aperture velocities are
systematically lower, although the actual difference cannot be
quantified.

\begin{table*}[t]
\begin{center}
\caption{Mean radial velocities of interstellar lines obtained from
\newsips\ SWP and LWP/LWR large aperture spectra}
\begin{tabular}{l c c c c c}
\hline
Target & v(SWP) & v(LW) & v(lit) & v(LW)-v(SWP) & v(lit)-v(LW)\\ &
km~s$^{-1}$ & km~s$^{-1}$ & km~s$^{-1}$ & km~s$^{-1}$ & km~s$^{-1}$ \\
\hline
RR~Tel		&  -69.5	&	-50.0	& -61.8		& 19.5		& -11.8	\\
$\zeta$ Oph	&  -24.7	&	-11.0	& -14.8		& 13.7		& -3.8	\\
BD+28~4211	&  -22.8	&	-3.8	& 		& 19.0		& 	\\
BD+75~325	&  -16.4	&	6.3	& 		& 22.7		& 	\\
HD~60753	&  18.8 	&	31.7	& 		& 12.9		& 	\\
HD~93521	&  -38.8	&	-20.1	& -10,-60	& 18.8		& 	\\
$\eta$ UMa	&		&	2.4	& -2.8		&		& -5.2	\\
$\lambda$ Lep 	&  		&	20.1	&  2,18		&		&       \\
$\zeta$ Cas	&		&	1.2	& -10.3		&		& -11.5	\\
\hline
\end{tabular}
\label{tab:rvel2}
\end{center}
\vspace{0.2cm}
Mean difference v(LW)-v(SWP)=17.7$\pm$3.7 km~s$^{-1}$ \\
Mean difference v(lit)-v(LW)=-8.0$\pm$4.2 km~s$^{-1}$ \\
\end{table*}

The reason for the discrepancy between the short and long
wavelength range velocity scales is not clear, while the
large/small aperture discrepancy in LW spectra is most likely
related to the transfer of the dispersion constants from the
small to the large aperture: the dispersion relations were
derived from spectra taken through the small aperture and then 
transferred to the large aperture on the basis of the assumed
aperture separations.

In order to provide an internally consistent wavelength scale
within the \ines\ system,  a velocity correction of  +17.7
km~s$^{-1}$ has been applied to the wavelength scale of SWP high
resolution spectra. The wavelength scale of LWP/LWR small
aperture spectra has been corrected by +13.7 km~s$^{-1}$. With
these corrections, the \ines\ velocity scale is  consistent with
the optical determinations.

\section{Conclusions}

\label{sec:summary}

In this paper we have discussed the overall quality of \iue\
high resolution spectra processed with the \newsips\ system.

The stability of the wavelength scale ($\approx$ 5 km~s$^{-1}$)
is  within the limits imposed by the acquisition and tracking
accuracy. No appreciable distortions in the wavelength scale
over the full spectral range or during the spacecraft lifetime
have been found.

A discrepancy of 9 km~s$^{-1}$ has been found in the velocities
derived from the two components of the Mg~II doublet at 2800
\AA\ in the LWP camera. The correct velocity is provided by the
K line measured on spectral order 83. No such discrepancy has
been found in LWR spectra.

The wavelength scales of \newsips\ short and long cameras
present  an  inconsistency,  which is well above the
repeatability errors quoted above. Measurements of narrow
interstellar lines have shown that SWP velocities are
systematically more negative by -17.7 km~s$^{-1}$, on average. A
similar discrepancy has been detected in long wavelength small
aperture spectra, whose velocities are more negative than those
from long wavelength large aperture spectra by -13.7
km~s$^{-1}$.

The determination of the inter--order background has greatly
improved with respect to the the \iuesips\ system, especially at
the shortest wavelengths, as shown by the absence of negative
fluxes in the cores of saturated absorption lines and by the
greater accuracy of equivalent width measurements.

The \ines\ system derives two spectra from each original
\newsips\ MXHI file. In the first one, the ``concatenated''
spectrum, the spectral orders are merged together and the
overlap regions are suppressed according to an algorithm which
computes suitable cut wavelengths which maximize the
signal-to-noise ratio. This spectrum contains an error vector
calibrated into absolute flux units, which is not available in
\newsips\ data.  Since the \ines\ high resolution spectra are
obtained from \newsips\ MXHI files, all previous considerations 
about the stability of the wavelength scale, the stability and
accuracy of the flux scale and the validity of the background
extraction are applicable to them. The second output product is
the ``rebinned'' spectrum, which is the ``concatenated''
spectrum after resampling at the low resolution wavelength step.

To correct for the discrepancies found in the \newsips\ high
resolution wavelength scale,  a velocity correction of +17.7
km~s$^{-1}$ for SWP spectra and of +13.5 km~s$^{-1}$ for LWP/LWR
small aperture spectra  has been applied to the \ines\
``concatenated'' spectra. With this correction,  the overall
\ines\ velocity scale is self--consistent, and agrees to  within
8 km~s$^{-1}$ with the optical velocity scale.

\begin{acknowledgements}

We would like to acknowledge the contribution  of all VILSPA
staff to the development and production of the \ines\ system, and
the referee, Dr. J.S. Nichols, for her useful comments.

\end{acknowledgements}


\begin{thebibliography}{}

\bibitem[{Barlow et al.}{1995}]{zetoph1}
Barlow, M.J., Crawford, I.A., Diego, F. et al., 1995, MNRAS 272, 333

\bibitem[{Brandt et al.}{1996}]{zetoph2}
Brandt, J.C., Heap, S.R., Beaver, E.A. et al., 1996, AJ 112,1128

\bibitem[{Cassatella et al.}{1999}]{ripple}
Cassatella, A., Altamore, A., Gonz\'alez-Riestra, R., Ponz, J.D., Barbero,
J., Wamsteker, W.,  1999, A\&AS, in press (Paper II)

\bibitem[{Frisch et al.}{1990}]{rvel1}
Frisch, P.C., Semback, K., York, D.G., 1990, ApJ 364, 540

\bibitem[{Garhart et al.}{1997}]{NEWSIPS2.0} 
Garhart, M.P., Smith, M.A., Levay, K.L., Thompson, R.W., 1997, IUE NEWSIPS
Information Manual, Version 2.0
 
\bibitem[{Morton}{1975}]{zetoph}
Morton, D.C., 1975, ApJ 197, 85

\bibitem[{Morton}{1991}]{lablines}
Morton, D. C., 1991, ApJS 77,119

\bibitem[{Nichols}{1998}]{joy98}
Nichols, J.S., 1998, in ``Ultraviolet Astronomy beyond the IUE Final
Archive'', ESA SP-413, eds. W. Wamsteker and R. Gonz\'alez-Riestra, p. 671

\bibitem[{Nichols \& Fesen}{1986}]{NF86} 
Nichols-Bohlin, J.S., Fesen, R.A., 1986, AJ 92, 642 

\bibitem[{Nichols \& Fesen}{1990}]{NF90} 
Nichols-Bohlin, J.S., Fesen, R.A., 1990, ApJ 353, 281

\bibitem[{Nichols \& Linsky}{1996}]{NL96} 
Nichols, J.S., Linsky, J.L., 1996, AJ 111, 517 

\bibitem[{pmr}{1999}]{RP99}
Rodr\'{\i}guez-Pascual, P.M., Gonz\'alez-Riestra, R., Schartel, N.,
Wamsteker, W., 1999, A\&AS 139, 183 (Paper I)

\bibitem[{Savage et al.}{1992}]{savage}
Savage, B.D., Cardelli, J.A., Sofia, U.J., 1992, ApJ 401, 706

\bibitem[{Smith}{1999}]{back}
Smith, M.A., 1999, PASP 760, 722

\bibitem[{Spitzer and Fitzpatrick}{1993}]{rvel3}
Spitzer, L., Fitzpatrick, E.L., 1993, ApJ 409, 299
 
\bibitem[{Tackeray}{1977}]{RRTel}
Tackeray, A.D., 1977, MNRAS 83,1 

\bibitem[{Vallerga et al}{1993}]{rvel2}
Vallerga, J.V., Vedder, P.W., Craig, N., Welsh, B.Y., 1993, ApJ 411, 729

\bibitem[{Wamsteker et al}{1999}]{ww} 
Wamsteker, W., Skillen, I., Ponz, J.D., de la Fuente, A.,
Barylak, M., Yurrita, I.,1999, Ap\&SS, in press

\bibitem[{Zuccolo et al.}{1997}]{zuc}
Zuccolo, R. Selvelli, P., Hack, M., 1997, A\&AS 124, 425

\end{thebibliography}
\end{document}